# A "Measure of Transaction Processing" 20 Years Later

Jim Gray
*Microsoft Research*

# A "Measure of Transaction Processing" 20 Years Later
Jim Gray
Microsoft Research
April 2005

"A Measure of Transaction Processing Power" [1] defined three performance benchmarks: *DebitCredit*: a test of the database and transaction system, *Sort*: a test of the OS and IO system, and *Copy*: a test of the file system.

*DebitCredit* morphed into TPC-A and then TPC-C. In 1985, systems were nearing 100 transactions per second, now they deliver 100,000 transactions per second, and a palmtop can deliver several thousand transactions per second (www.tpc.org, and [2]). Price-performance (measured in dollars/tps) has also improved dramatically as shown in Figure 1.

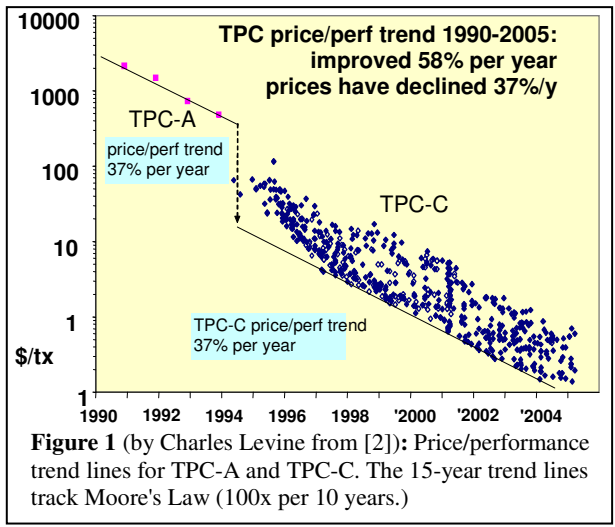

**Figure 1** (by Charles Levine from [2]): Price/performance trend lines for TPC-A and TPC-C. The 15-year trend lines track Moore's Law (100x per 10 years.)

The sort benchmark has seen similar improvements. The traditional Datamation "sort 1M records) now runs in a fraction of a second and has been replaced by PennySort (sort as much as you can for a penny), MinuteSort (sort as much as you can in a minute), and Terabyte sort (sort a trillion records). Each of these three benchmarks has a Daytona (commercial), and Indy (benchmark special) category [3].

This year Jim Wyllie of IBM Almaden Research won both Terabyte Sort and MinuteSort medals with his SCS (SAN Cluster Sort) [4] using SUSE Linux (SLES8) and IBM's General Parallel File System (GPFS) on an IBM 40-node 80-Itanium cluster with a SAN array of 2,520 disks. The machine delivered up to 14 GBps of IO bandwidth and sorted a terabyte in 437 seconds. SCS was able to sort 125 GB within a minute (wow!). It is hard to know the price of this system (now at UCSD) but the "list" price is at least $9M.

At the other end of the spectrum, Robert Ramey with his PostmansSort used a $950 Wintel box (3.2 GHz Pentium4, 2 Maxtor SATA disks, WindowsXP) to sort 16.3 GB in 979 seconds – setting a new Daytona Pennysort record [5].

Figure 2 shows that price-performance improved about 68%/year each year since 1985, handily beating Moore's 58%/year law. Sort speed (records sorted per second) doubled every year between 1985 and 2000. That doubling in part came from faster hardware, in part from better software, and in part from the use of LOTS more hardware (the year 2000 system used 1,962 processors and 2,168 disks.) In the last 5 years, peak sort speed has only improved 2.4x (about 20%/year improvement).

Performance improvements have been accomplished with multi-processors (hundreds of them). Price-performance improvements have come from cheaper and faster disks and from cheaper processors. But as Figure 3 shows, per-processor speeds seem to have plateaued. Sorted-records/second/processor (r/s/p for short) improved ten fold between 1985 and 1995. But speed has improved only 2.7 fold in the last decade.

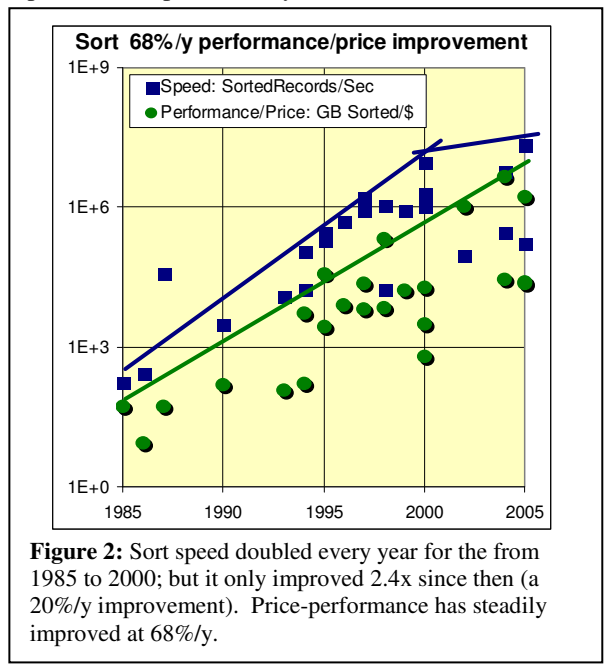

**Figure 2:** Sort speed doubled every year for the from 1985 to 2000; but it only improved 2.4x since then (a 20%/y improvement). Price-performance has steadily improved at 68%/y.

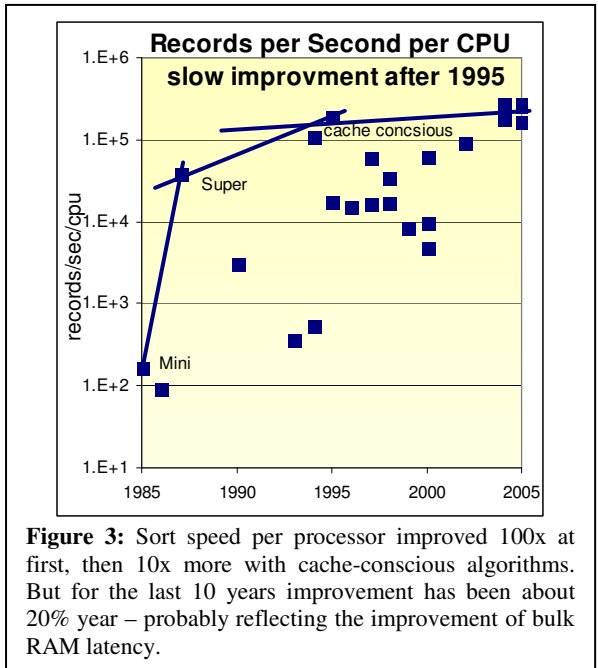

**Figure 3:** Sort speed per processor improved 100x at first, then 10x more with cache-conscious algorithms. But for the last 10 years improvement has been about 20% year – probably reflecting the improvement of bulk RAM latency.

The 1980's saw 200 r/s/p on a minicomputer to 38.5 k r/s/p on a Cray. In 1994, AlphaSort showed the importance of cache-conscious sorts and got to 111 k r/s/p. Since then, there has been a slow climb to 280 k r/s/p (e.g., Jim Willey's SCS Itanium TerabyteSort and the Pentium4 SkeenSort.)

I conjecture that this relatively slow improvement reflects the slow improvement in memory latency. All the algorithms are now cache conscious, so they are all limited by the speed of bulk memory – processor speed (and even cache speed) is not relevant here since sort cache misses are essentially random during the "comparison" and merge phase. But, that is just my guess. It would make an interesting study for the hardware architects -- speed problems may lie elsewhere. But, my guess is: Remember! It's the memory.

The tpcC benchmark defies this conjecture. Figure 4 plots the SQLserver throughput per cpu and per MHz for Intel/AMD processors (excluding Itanium) on the tpcC benchmark over a ten year period.[1]

It shows an initial steep 3-year learning curve of rapid improvement. Starting in 1998, tpmC per MHz began a gradual decline. In 2003 the introduction of AWE, much larger caches and multi-level caches, and then in 2004 the introduction of 64-bit addressing (with 64GB address spaces) allowed the systems more memory and brought the us back to about 10 tpmC per MHz.

---

[1] The following text was added after the article was published in the IEEE Data Engineering Bulletin.

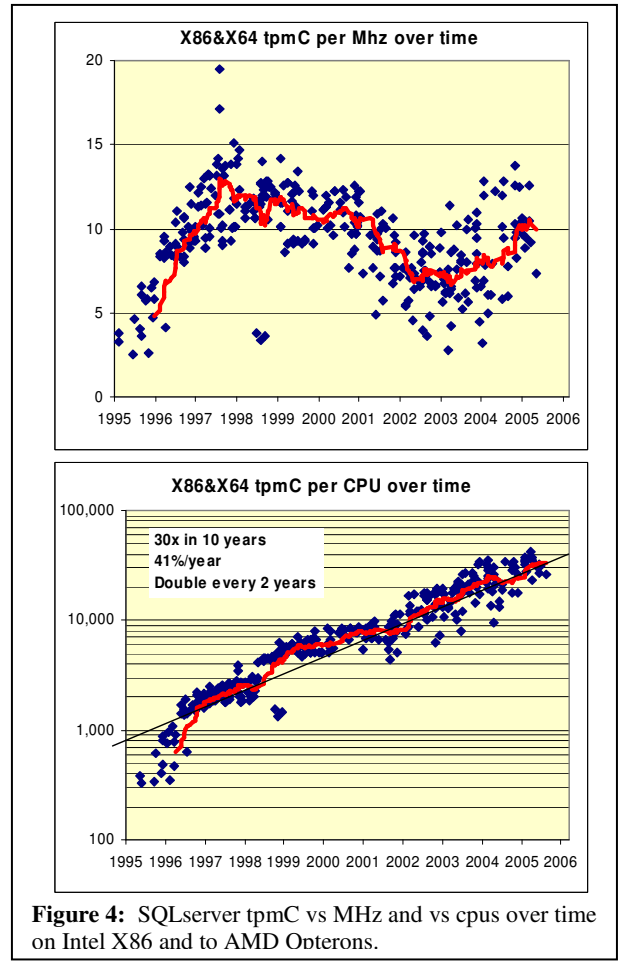

**Figure 4:** SQLserver tpmC vs MHz and vs cpus over time on Intel X86 and to AMD Opterons.

**Acknowledgments**
Charles Levine and Brad Waters were enormously helpful in building and understanding Figure4.